\documentclass[]{svjour3}
\usepackage{graphicx}
\usepackage{color}
\usepackage{amssymb}
\usepackage{amsmath}
\usepackage{mathptmx}      

\usepackage{latexsym}
\def\makeheadbox{\relax}
\begin{document}

\title{On identification of self-similar characteristics using the Tensor Train decomposition method with application to channel turbulence flow}
\titlerunning{Application of the Tensor Train decomposition to channel turbulence flow}

\author{Thomas von Larcher \and Rupert Klein}

\institute{Th. von Larcher \at Freie Universit\"at Berlin, Institute of Mathematics, Arnimallee 9, 14195 Berlin, Germany\\Tel.: +49-30-838-56093\\ \email{larcher@math.fu-berlin.de} \and R. Klein\at Freie Universit\"at Berlin, Institute of Mathematics, Arnimallee 6, 14195 Berlin, Germany
}

\date{} 

\maketitle

\begin{abstract}
A study on the application of the Tensor Train decomposition method to 3D direct numerical simulation data of channel turbulence flow is presented. 
The approach is validated with respect to compression rate and storage requirement. In tests with synthetic data, it is found that grid-aligned self-similar patterns are well captured, and also the application to non grid-aligned self-similarity yields satisfying results. It is observed that the shape of the input Tensor significantly affects the compression rate. 
Applied to data of channel turbulent flow, the Tensor Train format allows for surprisingly high compression rates whilst ensuring low relative errors.

\end{abstract}

\section{Introduction}
\vspace{-2pt}
Multidimensional data sets, data of dimension 3 or higher, require massive storage capacity that strongly depends on the (Tensor) dimension, {\it d}, and on the number of entities per dimension, {\it n}. The datasize or storage requirement scales with $\mathcal O (n^d), n=\max_{i} \{n_i\}$. It is that curse of dimensionality that makes it difficult to handle higher-order Tensors or big data in an appropriate manner. Tensor product decomposition methods, \cite{Hitchcock1927}, were originally developed to yield low-rank, i.e.\ data-sparse, representations or approximations of high-dimensional data in mathematical issues. It has been shown that those methods are as good as approximations by classical functions, e.g., like polynomials and wavelets, and that they allow very compact representations of large-data sets. Novel developments focus on hierarchical Tensor formats as e.g.\ the Tree-Tucker format, \cite{Hackbusch2009}, and the Tensor Train format, \cite{Oseledets2009}, \cite{Hackbusch2014}. Nowadays, those hierarchical methods are successfully applied in e.g., physics or chemistry, where they are used in many body problems and quantum states.

Here, we apply  the Tensor Train decomposition to data of a 3D direct numerical simulation (DNS) of a turbulence channel flow. We aim at capturing self-similar structures that might be hidden in the data as explained in the next paragraph. However, as those multiscale flow structures in highly irregular flows are not commonly aligned with the underlying grid but are translated, stretched, and rotated, we, at first, use synthetic data to evaluate the suitability of the method to generally detect self-similarity. 

In wall-bounded turbulent flows characteristic coherent patterns, heterogeneous structures limited in time appearing irregularly at varying positions, are observed in distinct spatial regions \cite{Kline1990}, \cite{Robinson1991}. A number of experimental and numerical studies were able to describe some (quasi-) coherent features rather accurately, e.g.\ quasi-streamwise vortical structures are identified as a form of quasi-coherent structures \cite{Bakewell1967}, \cite{Adrian2000}.  A primary focus of research is the near-wall region, i.e.\ $y^{+}<40$ ($y^{+}$ as wall coordinate), where low-speed streaks are observed moving away from the wall, and, consequently, a flow towards the wall is required for continuity reasons. Indeed, regions of such wall-directed rapidly moving fluid, called sweeps, are observed, e.g.\ \cite{Corino1969}, and it has been discussed that these patterns as well as the low-speed streaks positively contribute to the production of turbulent energy, e.g.\ \cite{Wallace1972}, \cite{Willmarth1972}. 

Today, in large eddy simulation (LES) it is still subject of intense research to model the unresolved so-called small scales. One branch follows the idea to reconstruct the fluctuations of the unresolved scales. For example, \cite{Adams2011} uses a stochastic approach of an adaptive deconvolution model for the subgrid scale closure; \cite{Hughes1995} suggests a variational finite element formulation; \cite{John2010} develops a variational multiscale method where adaption controls the influence of an eddy viscosity model. Self-similarity, however, is not resolved per-se in multiscale models. For this purpose, wavelet decomposition methods have been applied to data of turbulent flows, e.g.\ \cite{B07-FroehlichUhlmann2003}, and they were recently used to split the coherent signal from the incoherent part, e.g.~\cite{Khujadze2011}, \cite{Sakurai2016}. Similarity models, e.g. \cite{Liu1994}, and fractal models, e.g. \cite{Scotti1999}, are promising tools as they extrapolate self-similarity to the small scales.

Our study is concerned with the question of whether Tensor decomposition methods can support the development of improved understanding and quantitative characterisation of multiscale behavior of turbulent flows, cf.\ e.g.\cite{B07-TingEtAl2007}. Recent high-resolution numerical simulations obviously confirm established theoretic approaches that turbulent flows contain hierarchies of self-similar structures like vortex tubes or sheet-like flow patterns, e.g.\ \cite{Buerger2013}. Provided that our tests yield promising results, those quantitative features could be helpful in developing a LES closure approach based on and extending the idea of fractal or dynamic SGS models, e.g.\ \cite{B07-GermanoEtAl1991}, \cite{B07-Sagaut2006}. Therefore, if proved positively, a long-term goal would be the construction of a self-consistent closure for LES of turbulent flows that explicitly exploits the Tensor decomposition approach's capability of capturing self-similar structures. Our approach is automatically linked with the following questions: (i) Can real data from multiscale dynamics be approximated or represented by the Tensor Train decomposition technique and how compact are the resulting storage schemes, i.e what compression rate can be achieved at which level of accuracy? (ii) Does the Tensor Train approximated data retain the dynamics? (iii) Is the Tensor Train format suitable for detecting cascades of scales in real data and in turbulence data in particular? In this article, we present results of evaluating the Tensor Train format especially to (i). Promising results would encourage us for future work.

\section{The Tensor Train format}
\label{sec:TT}
In this section, we introduce the Tensor Train format, that is the Tensor decomposition method of choice for our study. A number of Tensor formats were developed in the past, e.g.\ the $r$-term approximation or so-called canonical format and the Tucker format \cite{Tucker1966},and we refer to e.g. \cite{Kolda2009} and \cite{Grasedyck2013} for surveys of low-rank approximation techniques and specifically Tensor approximation formats. Previous studies show that the Tensor Train format allows for a  much higher attainable compression rate compared to the low rank approximation formats just mentioned.

The Tensor Train format, however, is based on the Tucker decomposition. In the Tucker format a Tensor is decomposed into a product of a core Tensor and factor matrices. The factor matrices can be treated as the principal components of the Tensor and the core Tensor implies the level of interaction between the factor matrices. The Tucker format uses the advantages of the higher order singular value decomposition, hereafter HOSVD for short, with a closed set of low rank Tensors. The main disadvantage of the Tucker format is its high storage requirement as the core Tensor takes $r^{d}$ elements, $r$ as rank of the cores, resulting in a total storage requirement of $\mathcal{O}(dnr+r^{d})$. The Tensor Train format on the other hand is designed such that the storage requirement scales linearly with order $d$: $\mathcal{O}(dnr^{2})$, cf.~\cite{Oseledets2010}. 

The matrix SVD decomposes a matrix into a product of matrices, which then represents the original matrix: $\tens{A}=\tens{U}\,\tens{S}\,\tens{V^{T}}$, where $\tens{A}\in\mathbb{R}^{m\times n}$ is the original matrix, $\tens{U}\in\mathbb{R}^{m\times m}$ and $\tens{V}\in\mathbb{R}^{n\times n}$ are orthonormal matrices and $\tens{S}\in\mathbb{R}^{m\times n}$ is a diagonal matrix with the singular values, $\sigma_{n}$, of $\tens{A}$ on its diagonal. Note that $\sigma_{1}\ge\sigma_{2}\ge\ldots\ge\sigma_{n}\ge 0$ and the number of non zero singular values is equal to the rank of $\tens{A}$: $rank(\tens{A})=r$. If the spectrum of the singular values contains only a few large entries and a sharp cut-off to the remaining tail, considerable data compression rates can be realized by truncating the matrices to the significant part of the spectrum. E.g., an approximation of the original matrix, \tens{A}, with the first three singular values reads $\tens{A}\approx  \sigma_{1}u_{1}v_{1}^{T} + \sigma_{2}u_{2}v_{2}^{T} + \sigma_{3}u_{3}v_{3}^{T}$. Moreover, a truncation to rank $r$ results in a storage requirement of $mr+r+nr$ instead of $mn$ for the original matrix~$\tens{A}$.

\begin{figure}
	\centering
	\includegraphics{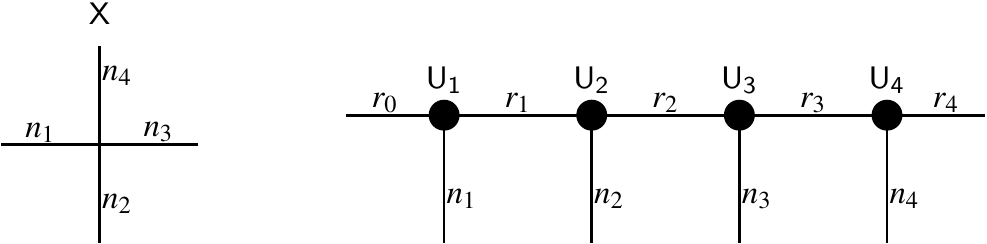}
	\caption{Decomposition of a 4th-order Tensor $\tens{X}(n_{1},n_{2},n_{3},n_{4})\in\mathbb{R}$ into the Tensor Train format. Left: sketch of the input Tensor. Right: sketch of the Tensor Train format with $\tens{U_{i}}$ ($i=1,\ldots,4)$ as the component Tensors $\tens{U_{1}}\in\mathbb{R}^{n_{1}\times r_{1}}$, $\tens{U_{i}}\in\mathbb{R}^{r_{i-1}\times n_{i}\times r_{i}}$, for $i={2,3}$ and $\tens{U_{4}}\in\mathbb{R}^{r_{3}\times n_{4}}$. Note that $r_{0}=r_{4}=1$ by definition.}
	\label{fig:Tensortrain}
\end{figure}

The key idea of Tensor Train decomposition is to separate a high-dimensional Tensor into its component Tensors (modes). The modes are then Tensors of order 3 by construction. 
Principally, the representation of a d-dimensional Tensor \tens{X} in the Tensor Train format reads
\begin{equation}
	\tens{X}(n_1,\ldots,n_d) = \sum_{k_1=1}^{r_1} \ldots \sum_{k_{d-1}=1}^{r_{d-1}} \tens{U_1}(n_1,k_1)\ \tens{U_2}(k_1,n_2,k_2)\ \ldots\ \tens{U_d}({k_{d-1},n_d}).
\end{equation}
The component Tensors $\tens{U_{i}}$ are obtained by the step-by-step application of the matrix SVD. The procedure is as follows: in the first step, the d-dimensional Tensor $\tens{X}$ is reshaped into a 2-dimensional matrix $\tens{A_{1}}\in\mathbb{R}^{n_{1}\times n_{2}\ldots n_{d}}$ to which a SVD is applied, $\tens{A_{1}}=\tens{U_{1}}\tens{S_{1}}\tens{V_{1}^{T}}$, and the first mode $\tens{U_{1}}\in\mathbb{R}^{n_{1}\times r_{1}}$ is obtained. The remaining matrices of the SVD are contracted to one matrix $\tens{A_{2}}=(\tens{S_{1}}\tens{V_{1}^{T}})\in\mathbb{R}^{r_{1}\times n_{2}\ldots n_{d}}$ which is used for the second step. In the second step, $\tens{A_{2}}$ is reshaped, at first, so that $\tens{A_{2}}\in\mathbb{R}^{r_{1}n_{2}\times n_{3}\ldots n_{d}}$. Then, a SVD is applied, again, and the second mode $\tens{U_{2}}\in\mathbb{R}^{r_{2}\times n_{3}}$ is obtained. By following this procedure successively, the remaining modes $\tens{U_{3}},\ldots,U_{d}$ are obtained. Note, that we need to apply the SVD $d$-times in total. As an example, figure~\ref{fig:Tensortrain} shows a graphical representation of the resulting Tensor Train network for an input Tensor of order 4.

The Tensor Train decomposition reveals some important topics: (a) in the Tensor Train network, the different modes $\tens{U_{i}}$ are linked by the ranks $r_{i}$ (as sketched in figure~\ref{fig:Tensortrain}). (b) the parameters $r_{i}$ determine the approximation quality. In case of Tensor approximation, as opposed to its representation, the $r_{i}$ are limited to some maximum rank, $r$, and a lower value of $r$ usually results in lower storage requirement for the approximated Tensor but also results in decreased approximation quality. The ranks $r_{k}$, therefore, are also called compression ranks or TT-ranks (TT as Tensor Train). This is because, in the sense of the procedure described previously, the value of $r$ determines the number of rows to be kept for the next step and sets the cut-off of the matrices $\tens{A_{i}}$. The hope is, that the omitted rows will be of small norm and the truncated component Tensors then will represent the significant part of the input Tensor. However, the optimal value of $r$ is not known a priori and it is not guaranteed that the omitted rows are indeed of small norm. That implies consequences for the application of the Tensor Train decomposition to real data as we will see later.
	
\section{Results}
\label{sec:results}
Here, we present the results of the Tensor Train approximation applied to data. We use the open source Tensor library {\it Xerus} developed at the Technical University Berlin, Germany, \cite{xerus}. The relative error of the approximation w.r.t. the original Tensor is measured in the Frobenius norm that, for a d-dimensional Tensor $\tens{X}$, reads
\begin{equation}
\vert\vert \tens{X} \vert\vert =\sqrt{\sum^{N_{1}}_{n_1=1} \sum^{N_{2}}_{n_2=1} \cdots \sum^{N_{d}}_{n_{d}=1} x^{2}_{n_{1}\cdots n_{d}}},
\end{equation} 
and the relative error in the Frobenius norm then reads
\begin{equation}
e = \frac{\vert\vert(\tens{U}-\tens{X})\vert\vert}{\vert\vert\tens{X}\vert\vert},
\end{equation}
with $\tens{U}$ as approximation of $\tens{X}$. 
\subsection{Principal operation of the Tensor Train approximation}
\begin{figure}
	\centering
	\includegraphics[width=0.8\textwidth]{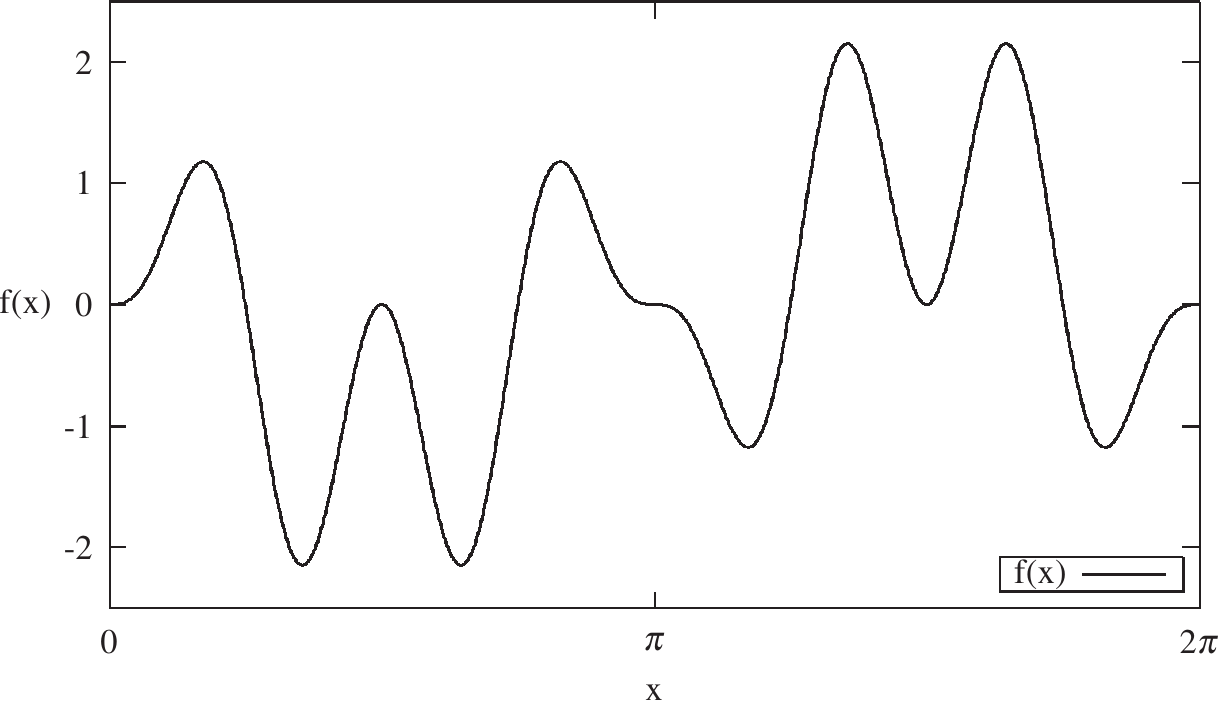}
	\vspace*{-4mm}
	\caption{Graph of $f(x)=3.125\sin(x)\sin(2x)\sin(4x),\ x\in[0,2\pi]$.}
	\label{fig:1}
\end{figure}

We compute a data series of a function, $f(x)$, that involves an overlay of sine-functions of different periods (figure~\ref{fig:1}). The function reads
\begin{equation}
	f(x)=3.125\sin(x)\sin(2x)\sin(4x),\qquad x\in[0,2\pi]
	\label{eq:1}
\end{equation}
Here, $f(x)$ is sampled with $2^{14}=16384$ digits. By construction, $f(x)$ involves similarity but not self-similarity in the proper sense.

Formally, the data series is a Tensor of order 1, i.e.\ a vector, and $\tens{X}(n_{1})\in\mathbb{R}$ with $n_{1}={1,\ldots,2^{14}}$. Without loss of generality, we can reshape the 1d-Tensor into a Tensor of order 3, that is $\tens{T}(n_{1},n_{2},n_{3})\in\mathbb{R}$ with $\tens{T}[2,2,4096]$. Note, that the choice of the shape of the Tensor $\tens{T}$ is initially arbitrary but has a significant impact on the result as we will see below.

\begin{figure}
	\centering
	\includegraphics[width=\textwidth]{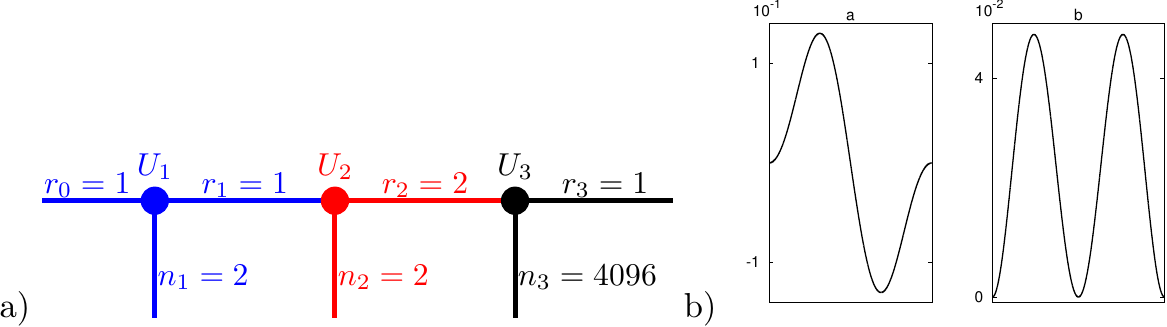}
	\caption{Left: schematic view of the Tensor, $\tens{T}[2,2,4096]$ written in the Tensor Train format. Note $r_{0}=r_{3}=1$ by definition. Colors denote the 3rd order Tensors $\tens{U_{1}}, \tens{U_{2}}, \tens{U_{3}}$. Right: the two basis vectors of $\tens{U_{3}}$.}
	\label{fig:2}
\end{figure}

Applying the Tensor Train decomposition to $\tens{T}$, it turns out that the input Tensor is approximated exactly, i.e.\ represented, for TT-rank $\ge 2$, and we find $r_{1}=1$ and $r_{2}=2$. Figure~\ref{fig:2}a shows a sketch of the approximated Tensor written in the Tensor Train format. The modes $\tens{U_{1}}$ and $\tens{U_{2}}$ read
\begin{equation*}
\tens{U_1}=\frac{1}{\sqrt{2}}\begin{pmatrix} -1 & 1 \end{pmatrix} \qquad \tens{U_2}=\frac{1}{\sqrt{2}} \begin{pmatrix} -1 & 1 \\ 1 & 1 \end{pmatrix},
\end{equation*}
and the basis is given by the components of mode $\tens{U_{3}}\in\mathbb{R}^{2\times4096}$ (see figure~\ref{fig:2}b). The storage of $\tens{T}$ in the Tensor Train format requires 8100 entries, which is less than half the storage requirement in the original Tensor notation. Note that, in the limit $r_{k}=1\ \forall k$, the Tensor approximation yields a relative error $e=0.33$ with a storage requirement of 4102 entries.

Now, we perform a test on the sensitivity of the Tensor Train approximation on the shape of the input Tensor. Suppose, we reshape the given data series, \eqref{eq:1}, into a Tensor $\tens{T_{2}}$ of order 4 where $\tens{T_{2}}[2,2,2,2048]$. Written in the Tensor Train format, TT-rank 4 is needed as minimum TT-rank to represent $\tens{T_{2}}$ (instead of TT-rank 2 for $\tens{T}$ in the former example). Note, that TT-rank 2 approximates the input Tensor with a relative error $e\approx0.05$. Table~\ref{tab:1} shows the relative error and the storage requirement for TT-rank 1 to 4.   

\begin{table}[t]
	\caption{Relative error, $e$, and storage requirement for approximation of Tensor $\tens{T_{2}}$ for TT-ranks 1 to 4.}
	\centering
	\label{tab:1}
	\begin{tabular}{lll}
		\hline\noalign{\smallskip}
		TT-rank & $e$ & storage requirement \\[3pt]
		1 & 0.448 & 2056 \\
		2 & 0.047 & 4112 \\
		3 & 0.004 & 6164 \\
		4 & 1.8e-15 & 8216 \\
		\noalign{\smallskip}\hline
		\end{tabular}
\end{table}

\subsection{Detection of self-similar structures}
\begin{figure}
	\centering
	\includegraphics[width=0.8\textwidth]{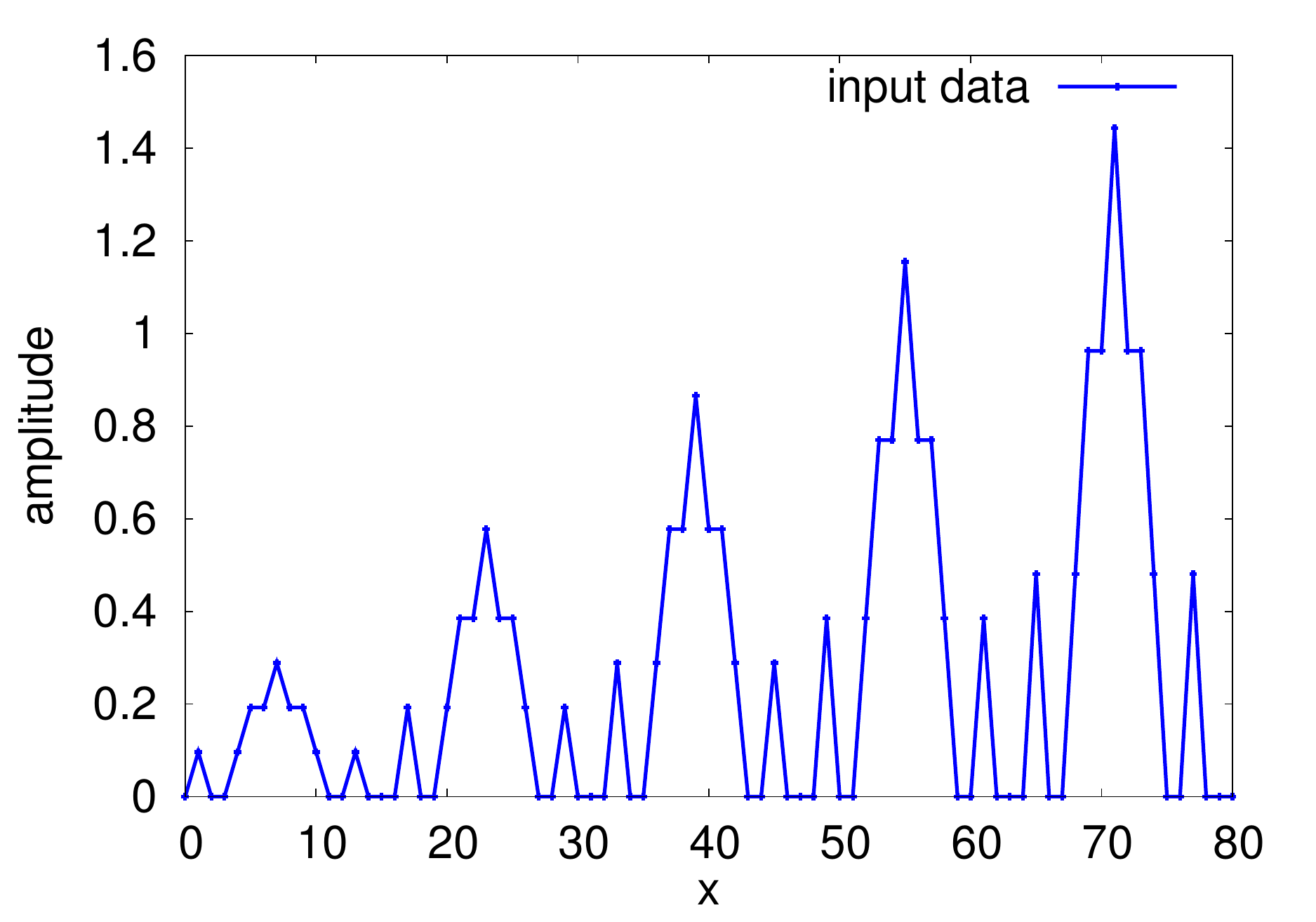}
	\caption{Sequence of the self-similar triangular data series. The data are repeated and scaled-up every 16\,digits up to 1024 \,digits in total.}
	\label{fig:4}
\end{figure}

Now, we extent the first example and test the Tensor Train format to detect self-similar structures hidden in data series. For this purpose, we compute a data series that consists of a $2^{4}=16$\,digit-sequence of triangles. Thus, it is in line with the power of two sequence. The 16 digit sequence is not only repeated but is also scaled-up in every loop, see figure~\ref{fig:4}. The data series comprises $2^{10}=1024$ digits in total. As shown previously, the impact of the shape of the input Tensor on the resulting storage requirement can be significant. Therefore, we reshape the data series in a Tensor of order 9 where each of the 9 dimensions have 2 entries except the second to last one that has 4 entries, i.e. $\tens{T}_{3}[2,2,2,2,2,2,2,4,2]$. Thus, the self-similar structure in the data series is maintained and, furthermore, remains grid-aligned and is not disturbed by splitting of the Tensor. We find TT-rank 2 as minimum rank to represent the input Tensor in the Tensor Train format. The storage requirement, 66 entries ($\approx6.4$\,\% of the original Tensor notation), is very low. Variations of the input Tensor's shape results in very low storage requirements, too, but it turns out that the shape of $\tens{T}_{3}$ demands the lowest storage requirement and the lowest TT-rank. Table~\ref{tab:2} shows results for varieties of the input Tensor entries.

 \begin{table}
\caption{Self-similar triangles data series: minimum TT-ranks and storage requirement to represent different shapes of the input Tensor $\tens{T}_{3}$. Note that $\tens{T}$ is a Tensor of order 9 in all cases.}
\label{tab:2}
\begin{tabular}{lll}
\hline\noalign{\smallskip}
$\tens{T}_{3}$ & TT-ranks ($r_{1}, \ldots,r_{8}$) & storage requirement \\[3pt]
$[2,2,2,2,2,2,2,2,4]$ & 2,2,2,2,2,1,2,3 & 70 \\
$[2,2,2,2,2,2,2,4,2]$ & 2,2,2,2,2,1,2,2 & 66 \\
$[2,2,2,2,2,2,4,2,2]$ & 2,2,2,2,2,1,3,2 & 70 \\
$[2,2,2,2,2,4,2,2,2]$ & 2,2,2,2,2,2,3,2 & 82 \\
$[2,2,2,2,4,2,2,2,2]$ & 2,2,2,2,1,2,3,2 & 70 \\
\noalign{\smallskip}\hline
\end{tabular}
\end{table}

\begin{figure}
	\includegraphics[width=\textwidth]{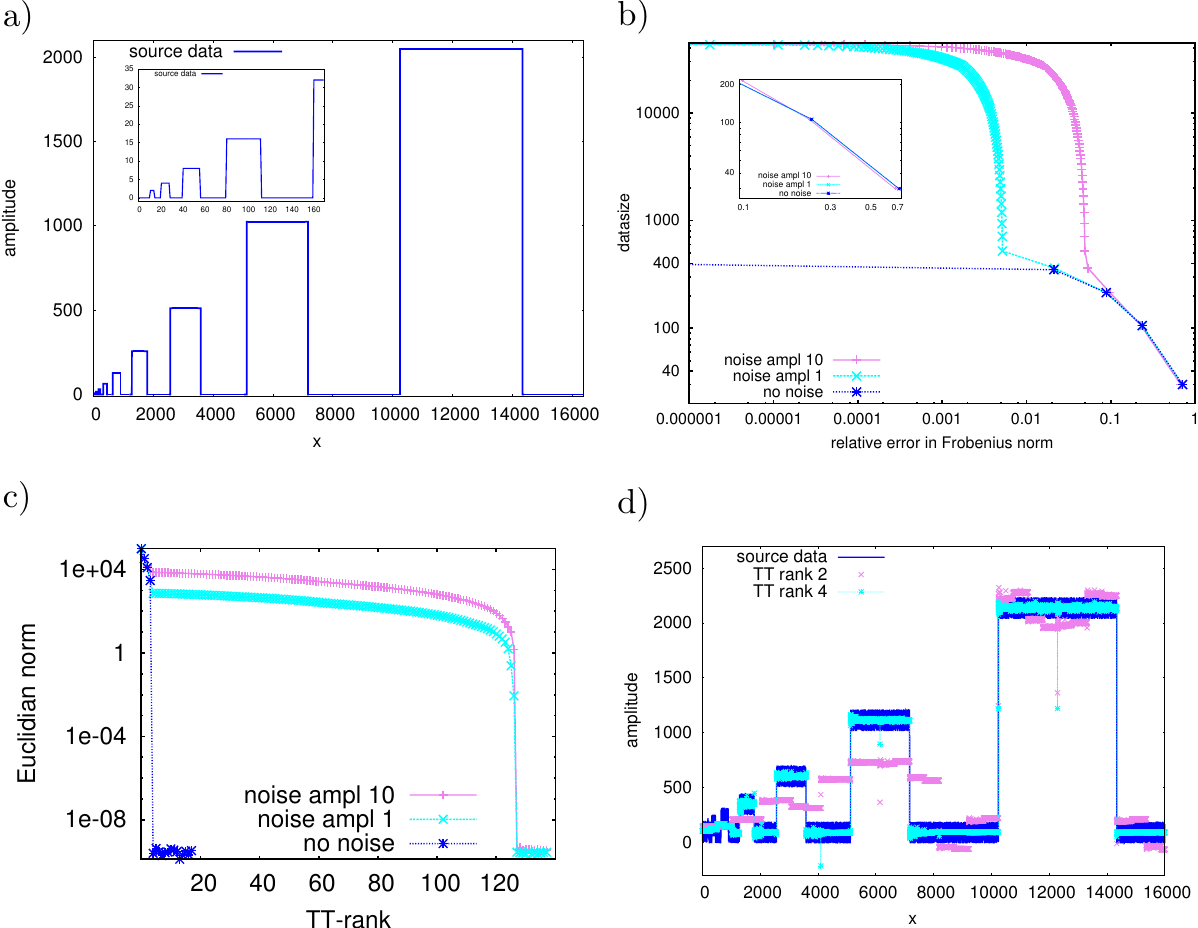}
	\caption{Test of the top-hat function. a) synthetic top-hat data series, noise amplitude 0. b) datasize over relative error for different TT-ranks: no noise TT-rank $1,\ldots,4$ (blue, dotted line marked by star signs); noise factor 1 (cyan, dashed line marked by crosses) and noise factor 10 (magenta, solid line marked by plus signs) TT-rank $1,2,\ldots,126$. Note that Tensor approximation with TT-rank 1 has the lowest storage requirement but the largest error. c) Euclidian norm of the difference between original and approximated data series over TT-ranks. d) approximated data series with noise amplitude 10 for TT-rank 2 (magenta, marked by crosses) and TT-rank 4 (cyan, marked by star signs).}
	\label{fig:5}
\end{figure}

Finally, we consider the properties of the Tensor Train format in case the input Tensor involves self-similar structures not aligned with the grid, that is, the self-similar patterns are not in line with the power of two sequence. For that purpose we compute a top-hat function where the squares are scaled up, see figure~\ref{fig:5}a. The data series contains $2^{14}=16384$ entries in total. To test the effect of noisy data, we perform two additional tests and add random noise of different amplitude, here amplitude factor 1 and 10, to the data series. We reshape the data series into a Tensor of order 14, thus each dimension has 2 entries, $\tens{T}_{4}[2,2,2,2,2,2,2,2,2,2,2,2,2,2]$. With this arbitrary choice, we intentionally neglect our a priori knowledge on the character of the data series. This is reasonable as we can not expect detailed knowledge of the data structure in case of real data.

Figure~\ref{fig:5}b shows the storage requirement (hereafter denoted as datasize) over the relative error for both, the data series without noise and with noise amplitude factor 1 and 10, for a number of TT-ranks. In case of no noise Tensor $\tens{T}_{4}$ is represented with TT-rank 5 where relative error is about $10^{-15}$, and storage in the Tensor Train format requires about 500 entries thus $\approx3.0$\,\% of the original datasize. Limiting the TT-rank to 2 (4) yields approximation error of about 0.2 (0.02)\,\%, and storage of the approximated Tensor in the Tensor Train format requires about 100 (350) entries, i.e.\ 0.6 (2.1)\,\% of the datasize in original Tensor notation. Approximation of the noisy data series for TT-rank 5 to about 80 shows a drastic increase in the storage requirement whilst only small changes in the relative error. Obviously, this is due to the small-scale noise the approximation of which naturally requires many more data entries.

Both data series, with noise amplitude 1 and with noise amplitude 10, are approximated exactly at TT-rank 128 at which the relative error drops to $10^{-15}$. In both cases, the storage requirement of the representations in the Tensor Train format (43690 entries for both cases) substantially exceeds the storage requirement of the source data in the original Tensor format (16384 entries). For both noisy data series, we find the original datasize at TT-rank 45 where $e\approx0.03$ (noise factor 10) and $e\approx0.003$ (noise factor 1), resp. Furthermore, for a given datasize, the relative error strongly depends on the noise amplitude as long as the noise is not well approximated.

Interestingly, the relative error agrees well with that of the no-noise case up to TT-rank 4, TT-rank 3 for noise amplitude 10, indicating that the self-similar top-hat structure, large-scaled relative to the small scale noise, but not the noise itself is represented by low TT-ranks. Figure~\ref{fig:5}c shows the euclidian norm of the difference between the original Tensor and the approximated Tensor. At low TT-ranks, the euclidian norm agrees well in all cases underlining the previous statement that large-scale structures are resolved at low TT-ranks. At large TT-ranks, the graphs of both noisy data series show a significant drop at TT-rank 127 specifying the exact approximation of the input Tensor.  

Figure~\ref{fig:5}d confirms that picture since the self-similar top-hat structure but not the noisy part (small-scaled w.r.t. the top-hat structure) is approximated well at TT-rank 4 (except the small-scale top-hats at the beginning of the data series). With TT-rank 2 as maximum permitted TT-rank, the approximation of the top-hat structures is less accurate but the general trend is obtained, at least. That result is of course linked with the mathematics of the algorithm discussed previously since the given TT-rank is linked with the number of rows, i.e.\ number of singular values, to be kept in the Tensor Train decomposition procedure (cf.\ section~\ref{sec:TT}). Low rank, therefore, means a loss of information particularly about small-scales, that is usually hidden in higher eigenvalues of the matrices. In other words, large-scale structures are already being represented whilst small-scale patterns are still coarsely approximated.

In summary, the tests with different synthetic data series show that approximation in the Tensor Train format is efficient and promising in detecting self-similar patterns in particular if those structures are grid-aligned. Even if the self-similar structures are not grid-aligned, the approximation in the Tensor Train format yields satisfying results particularly for data series without noise. Furthermore, we found that the minimal TT-rank to represent a data series in the Tensor Train format is sensitive to the shape of the input Tensor. The latter statement has been also discussed by \cite{Ballani2014}.

\subsection{Channel turbulence flow}
\begin{figure}
	\includegraphics[width=\textwidth]{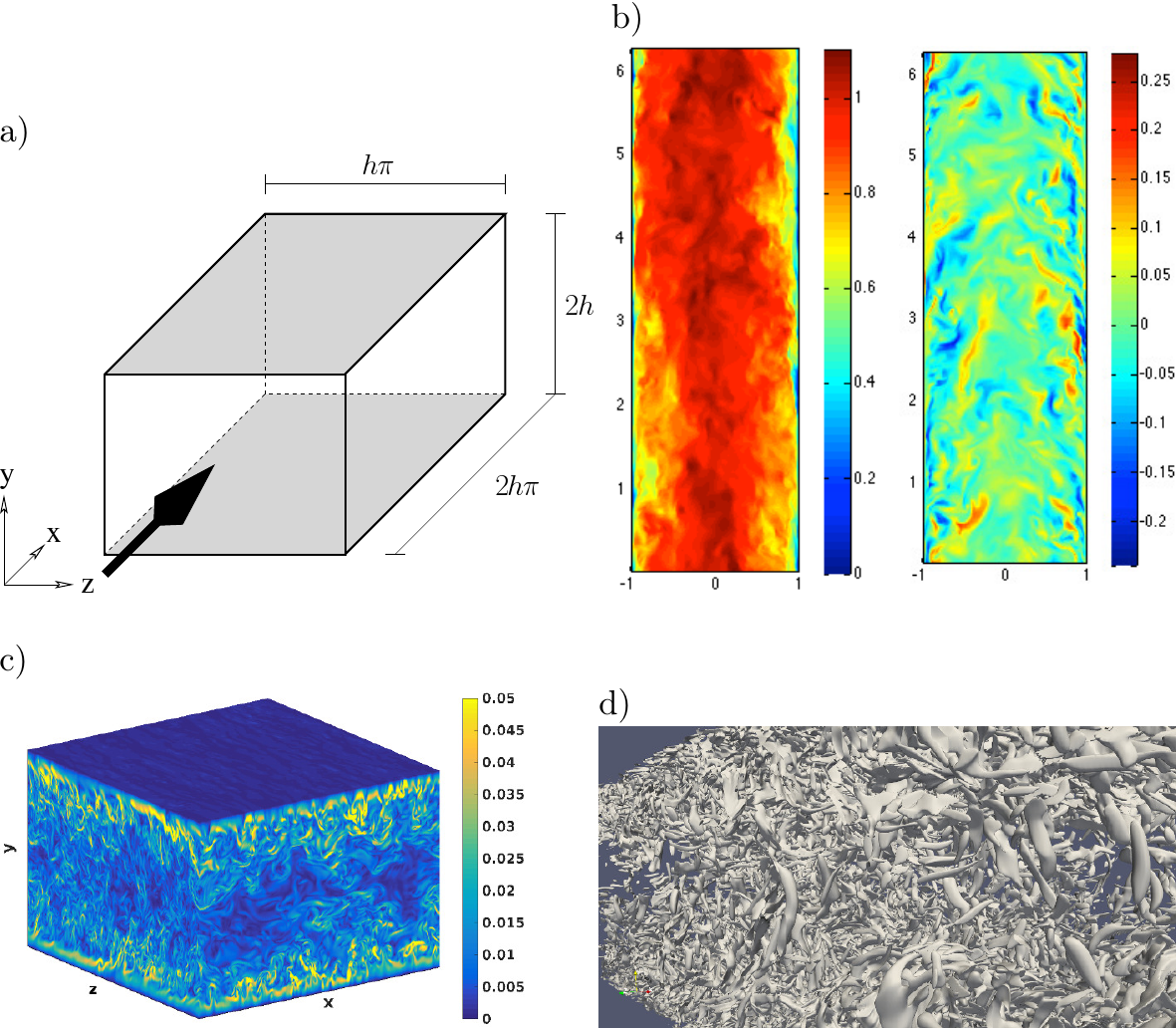}
	\caption{Turbulent channel flow. a) sketch of the channel geometry (courtesy of M.\ Waidmann) with $x$-dimension set to $2h\pi$, $z$-dimension to $h\pi$ and $y$-dimension to $2h$, with $h=1$ for the computations. b) ($x,y$)-slice of the ($x$)-component (left panel) and of the $z$-component of the velocity (right panel). Note that the main flow direction is from bottom to top. c) slices of the vorticity magnitude at the boundaries of the channel geometry, d) iso-surface map of the q-criterion.}
	\label{fig:6}
\end{figure}

We use data of 3D direct numerical simulation (DNS) - by using a pseudo-spectral Fourier-Chebyshev method- of an incompressible, isothermal plane channel flow, figure~\ref{fig:6}a,  computed by \cite{Uhlmann2000b}, \cite{Uhlmann2000a}, to which we refer for an in-depth description of data generation. 

Data are generated at $Re_{\tau}=590$, where $Re_{\tau}$ is the friction-based Reynolds number. The channel geometric factor $h$ has been set to $h=1$.
The original grid spatial resolution is $600\times 385\times 600$ in ($x$,$y$,$z$). Here, we make not use of data of the original grid but of the so-called {\sl fine grid} that is $600\times 352\times 600$ in ($x$,$y$,$z$), and we have applied a post-processing procedure to obtain a constant grid increment throughout the $y$-direction instead of the polynomial distribution of the original grid.  We refer to \cite{Larcher2015} for a detailed description of the post-processing algorithms that convert the original data to the equidistant fine grid data.

In wall units the grid increment of the fine grid is $\Delta y^{+}=3.37$. Compared to the signatures of the original grid data, e.g.\ fig. 3 and 4 in \cite{Uhlmann2000c}, no quantitative loss of information is recognized due to the fine grid post-processing procedure. However, the resolution of the near-all region is significantly reduced. In particular, the viscous sublayer ($y^{+}<5$) is not resolved anymore.

The scalar q-criterion proposed by \cite{Hunt1987}, \cite{Hunt1988} is a measure to demarcate vortex structures within turbulent flows. The q-criterion reads
\begin{equation}
q = \frac{1}{2} (\vert\vert A\vert\vert^2 - \vert\vert S\vert\vert^2),
\end{equation}
where $\vert\vert S\vert\vert = [tr(SS^t)]^{1/2}$, $\vert\vert A\vert\vert = [tr(AA^t)]^{1/2}$, with $S$ and $A$ as the symmetric and anti-symmetric  part of the velocity gradient Tensor $\nabla U$, resp. Figure~\ref{fig:6}d shows an iso-surface map of the q-criterion for a given snapshot. Here, vortex tubes of different size and shape, rotated and stretched, can be identified indicating the highly turbulent flow.

Here, we apply the Tensor Train decomposition to velocity data (fig.~\ref{fig:6}b), the datasize is $380160000\approx 10^{9}$ entries per snapshot. 

Since the tests with synthetic data series already have revealed the sensitivity to the shape of the input Tensor, see the previous subsection, we consider two different approaches. Firstly, the input Tensor retains the original data structure, that is, it is a 4th-order Tensor $\tens{T}[600,352,600,3]$ where the last dimension contains the components of velocity. This approach is referred to as TT-approximation hereafter. On the other hand, we decompose every dimension into its prime factors, resulting in a Tensor of order 19
\begin{equation}
 	\tens{T}[n_{1},\ldots,n_{19}]=\tens{T}[2,2,2,3,5,5,\ 2,2,2,2,2,11,\ 2,2,2,3,5,5,\ 3].
	\label{eq:19d}
 \end{equation}
 The latter approach is referred to as QTT-approximation hereafter as it is similar to the Quantics Tensor Train (QTT) approach, \cite{Khoromskij2011}, \cite{Oseledets2010a}, inasmuch as the binary representation is realized as well as possible.

\begin{figure}
	\includegraphics[width=0.8\textwidth]{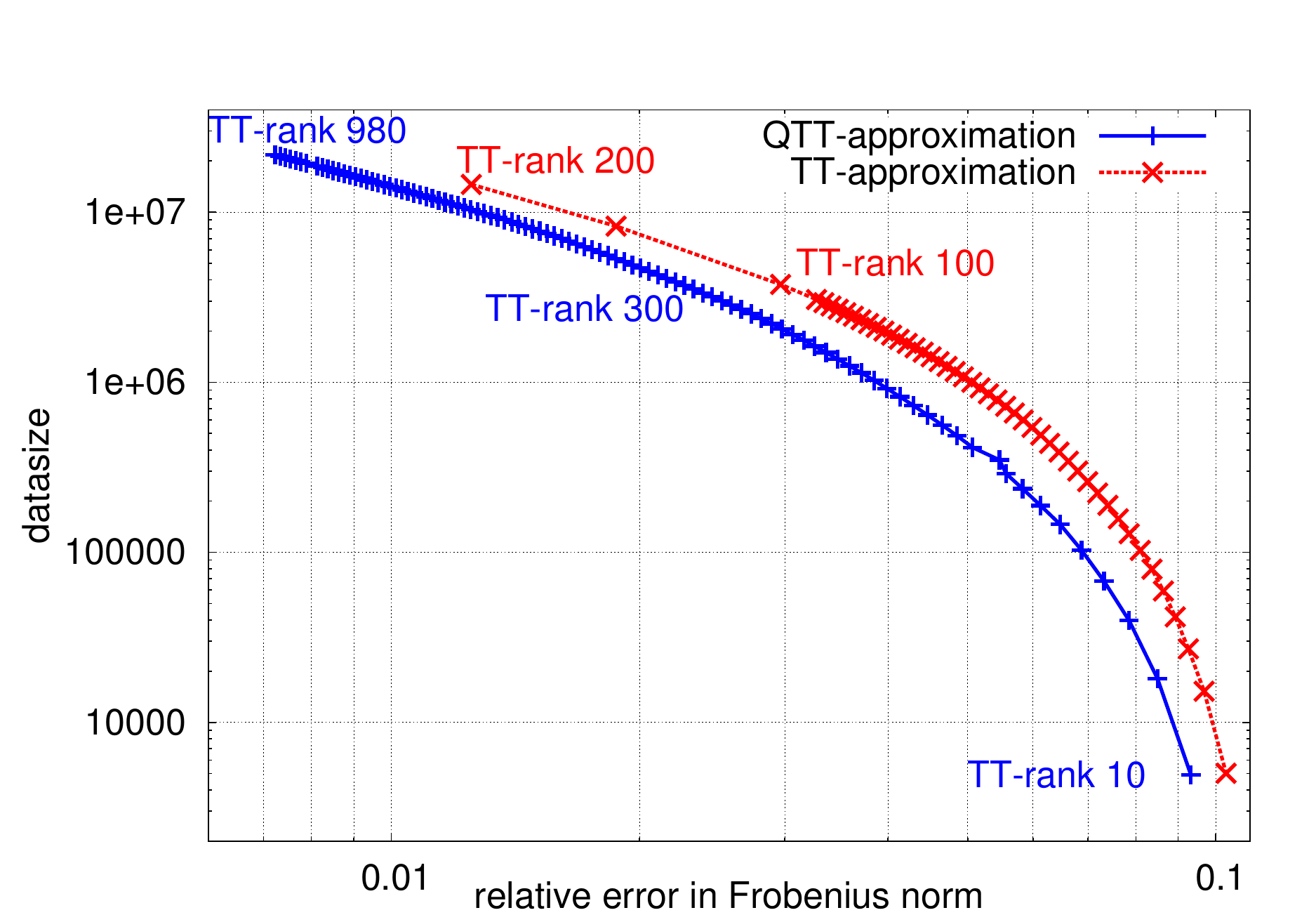}
	\caption{Approximation of velocity data. Storage requirement over relative error, $e$, for QTT- (blue, solid line marked by plus signs) and TT-approximation (red, dotted line marked by crosses) for various TT-ranks. Note, that we set different TT-ranks in the QTT- and TT-approximation: In the TT-approximation, the increment is set to 2 up to TT-rank 100, and TT-rank 150 and 200 is also shown. In the QTT-approximation, the increment is set to 10 up to TT-rank 980.}
	\label{fig:7}
\end{figure}

Figure~\ref{fig:7} shows the results of the Tensor Train decomposition for both approaches. Here, the diagram shows storage requirement over relative error for various TT-ranks.  Obviously, the QTT-approximation yields less relative error with significantly less storage requirement. For example, given a relative error of 0.085 the QTT-approximation demands a storage requirement of about 18000 entries but TT-approximation demands about 60000 entries. Induced by the different shapes of the input Tensors the necessary TT-rank is much larger in the QTT-approach.

\begin{figure}
	\includegraphics[height=\textwidth]{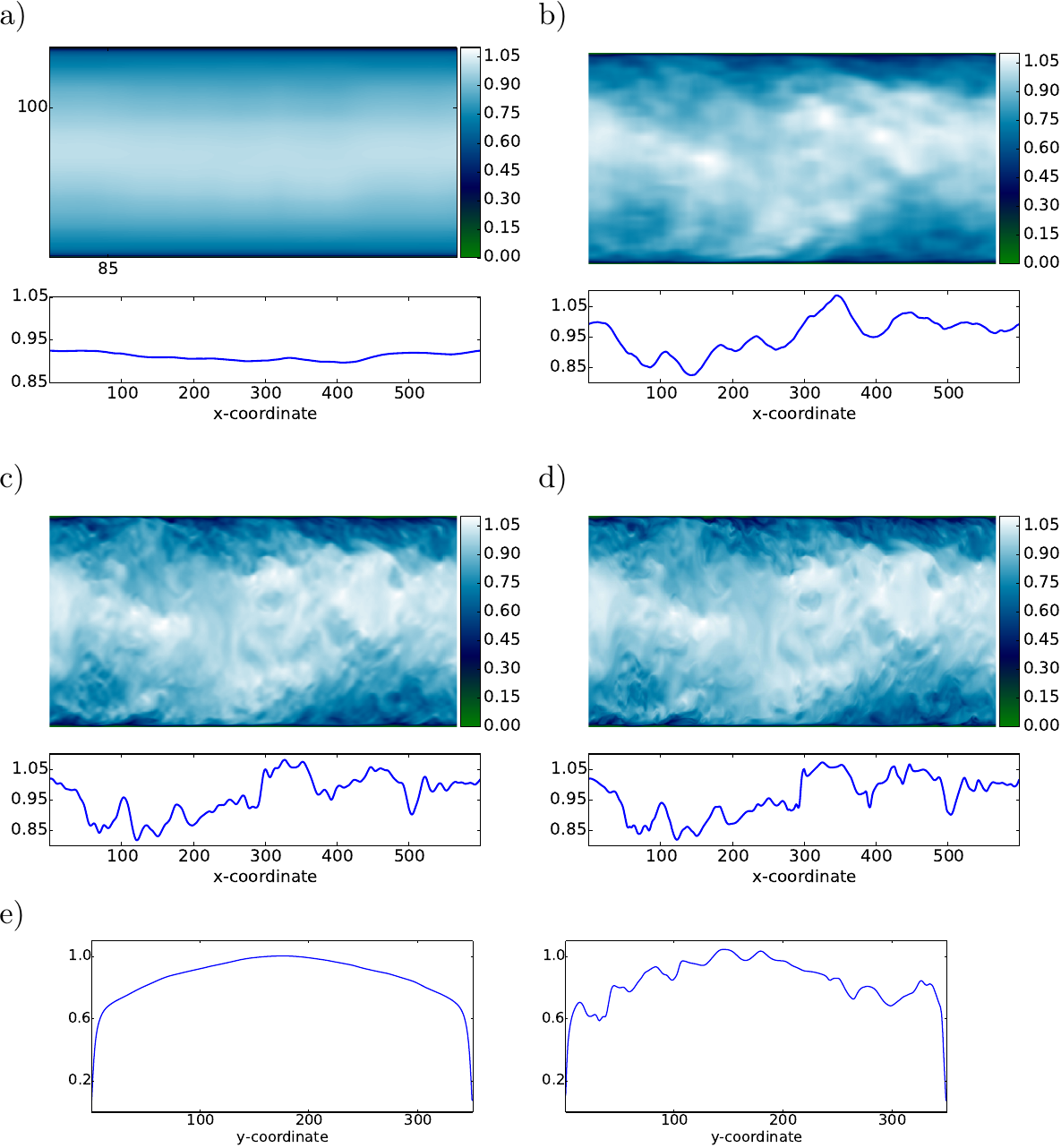}
	\caption{TT-approximation of the velocity field in the Tensor Train format. ($x,y$)-slice of velocity magnitude at $z=300$ (mid-channel) and cross-section along the $x$-coordinate extracted at $(y,z)=(100,300)$. a) for TT-rank 2  ($e\approx0.103$), b) for TT-rank 22  ($e\approx0.072$), c) for TT-rank 100  ($e\approx0.030$), d) original data; e) cross-section along the $y$-coordinate  extracted at $(x,z)=(85,300)$; TT-rank 2 (left panel) and original data (right panel). The appropriate coordinate $x=85$ as well as $y=100$ is marked in panel a).}
	\label{fig:8}
\end{figure}

\begin{figure}
	\includegraphics[width=\textwidth]{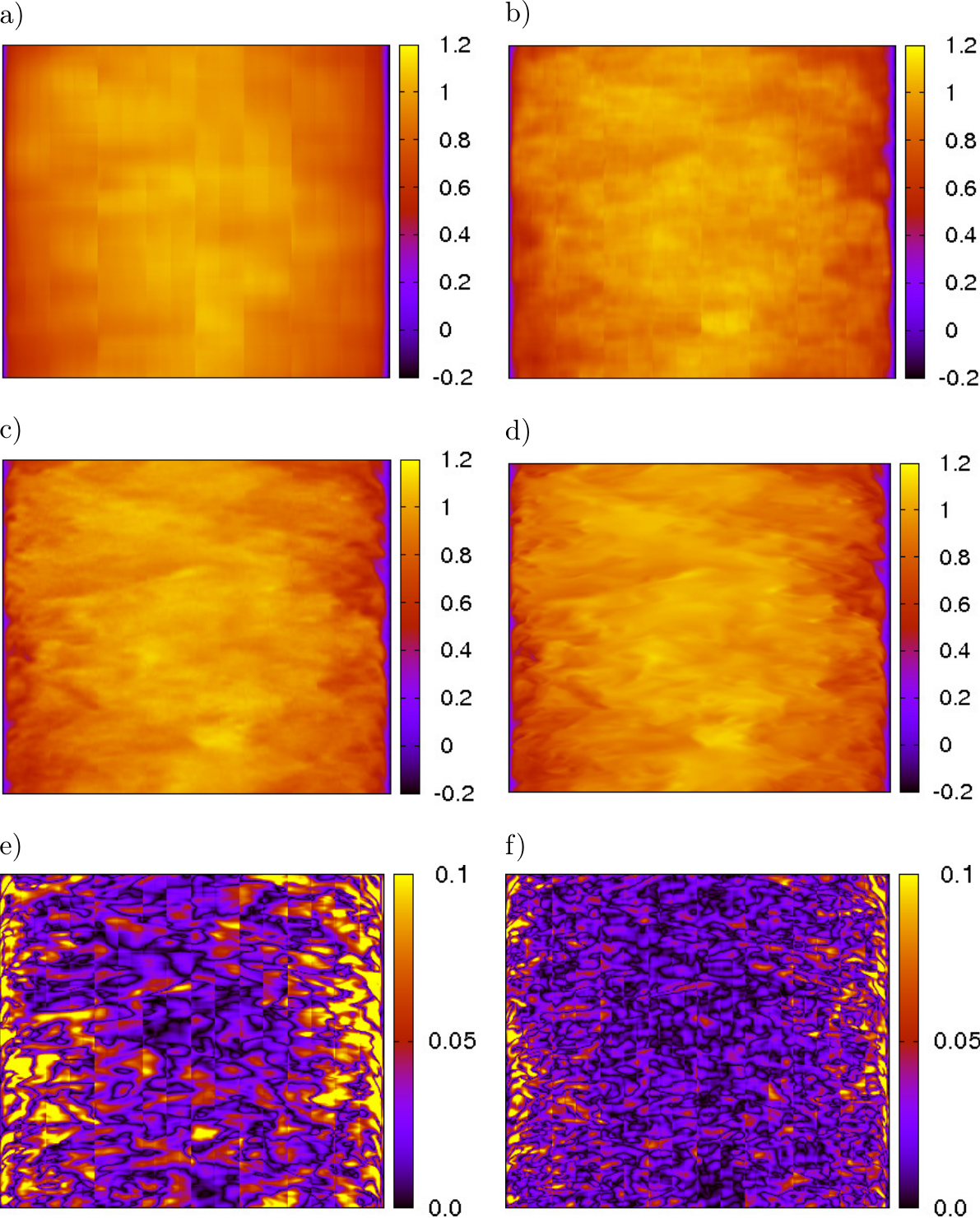}
	\caption{QTT-Approximation of the velocity field. ($x,y$)-slice of $x$-component of velocity at $z=300$ (mid-channel). a) for TT-rank 20  ($e\approx0.085$), b) for TT-rank 100  ($e\approx0.055$), c) for TT-rank 420 ($e\approx0.020$), d) original data; e) difference between approximated data and original data of the x-component of velocity for TT-rank 20, f) as e) but for TT-rank 100.}
	\label{fig:9}
\end{figure}

Figures~\ref{fig:8} for the TT-approximation and figure~\ref{fig:9} for the QTT-approximation displays 2D-slices of the approximated data for selected TT-ranks, and also a 2D-slice of the original data is shown for comparison. Figure~\ref{fig:8} displays the magnitude of velocity computed from the components after TT-approximation and figure~\ref{fig:9} shows the streamwise component of the velocity. In addition, 1D cross-sections along $x$-coordinate are given in figure~\ref{fig:8}. Generally, images of lower TT-ranks show a smooth reflection of the turbulent flow, and with increasing TT-rank small-scale patterns are mapped with enhanced resolution. Interestingly, in the TT-approximation, TT-rank 2 involves the box-shaped mean velocity profile typical for channel turbulence flows, figure~\ref{fig:8}e, cf. e.g.~\cite{Pope2000}. In that sense, Tensor Train decomposition restricted to TT-rank 2 apparently acts as an averaging process.

Now, we are interested in identifying those regions in the channel flow that show the most significant deviations from the original in the approximated data. In both cases, TT-approximation and QTT-approximation, reconstruction of the velocity field for higher TT-ranks, for example TT-rank 100 in figure~\ref{fig:8}c and \ref{fig:9}b, already shows several details of small-scale features of the turbulent flow by eye, even if the relative error is significant. For the QTT-approximation, figure~\ref{fig:9}e for TT-rank 20 and \ref{fig:9}f for TT-rank 100 shows the difference between the approximated and the original data. Obviously, significant deviations are located towards the near-wall region and the flow interior is mostly unaffected. This picture is becoming more and more visible the more details of the small-scale features are resolved, thus with increasing TT-rank.
 

\section{Concluding remarks}
\label{sec:conclude}
In this article, the Tensor Train decomposition technique, a specific branch of the family of Tensor product decomposition methods, has been applied to DNS data of channel turbulence flow. We showed that the Tensor Train format yields significant compression rates whilst ensuring low relative errors. In particular, the QTT-approximation, that in our definition is decomposing the entries of each dimension of the input Tensor into its prime factors, shows considerably better results than the approximation of the original data Tensor of order 4.

Applied to channel turbulence data, the Tensor Train format allows for surprisingly high compression rates. QTT-approximation at TT-rank 100 results in a compression factor of roughly 1000 and in a relative error of about 5\,\%. At TT-rank 2000, the relative error is about 0.2\,\% with a compression factor of about 5, that is a storage requirement of $\approx10^{8} $ (instead of $\approx10^{9}$ for the original data). However, it demonstrates that low-rank representation of those highly irregular and chaotic flows in the Tensor Train format can not be expected.

Our study is concerned with the detection of and quantitative characterization of self-similar patterns in turbulent flow data. Therefore, we, firstly, applied synthetic data to demonstrate the Tensor Train decomposition´s capability to capture self-similarities. The synthetic data tests with and without noise provide promising results. Grid aligned self-similarity is well captured and non grid-aligned self-similarity is approximated at low TT-ranks strengthening the suitability of the method. Moreover, with no noise, non grid-aligned self-similar structures are approximated exactly at low TT-ranks.

\begin{acknowledgements}
This research has been funded by Deutsche Forschungsgemeinschaft (DFG) through grant CRC 1114  'Scaling Cascades in Complex Systems', Project B04 'Multiscale Tensor decomposition methods for partial differential equations'. The authors thank Prof. Illia Horenko (CRC 1114 Mercator Fellow) as well as Prof. Reinhold Schneider and Prof. Harry Yserentant for rich discussions and for steady support. Thomas von Larcher thanks Sebastian Wolf and Benjamin Huber (both at TU Berlin, Germany) very much for developing the Tensor library \textsl{xerus} which has been used for data analysis, as well as for their round the clock support in the project. The data were generated and processed using resources of the North-German Supercomputing Alliance (HLRN), Germany, and of the Department of Mathematics and Computer Science, Freie Universit\"at Berlin, Germany. The authors thank Alexander Kuhn and Christian Hege (both at Zuse Institute Berlin, Germany) for steady support in data processing and data visualisation. 
\end{acknowledgements}

\bibliographystyle{spmpsci} 
\bibliography{tensor-bibfile} 

\end{document}